\documentclass[12pt]{article}
\usepackage{amssymb}
\usepackage{epsfig}
\voffset= - 0.3in
\hoffset= - 0.7in         

\catcode`\@=11
\def\marginnote#1{}

\newcount\hour
\newcount\minute
\newtoks\amorpm
\hour=\time\divide\hour by60
\minute=\time{\multiply\hour by60 \global\advance\minute by-\hour}
\edef\standardtime{{\ifnum\hour<12 \global\amorpm={am}%
        \else\global\amorpm={pm}\advance\hour by-12 \fi
        \ifnum\hour=0 \hour=12 \fi
        \number\hour:\ifnum\minute<10 0\fi\number\minute\the\amorpm}}
\edef\militarytime{\number\hour:\ifnum\minute<10 0\fi\number\minute}

\def\draftlabel#1{{\@bsphack\if@filesw {\let\thepage\relax
   \xdef\@gtempa{\write\@auxout{\string
      \newlabel{#1}{{\@currentlabel}{\thepage}}}}}\@gtempa
   \if@nobreak \ifvmode\nobreak\fi\fi\fi\@esphack}
        \gdef\@eqnlabel{#1}}
\def\@eqnlabel{}
\def\@vacuum{}
\def\draftmarginnote#1{\marginpar{\raggedright\scriptsize\tt#1}}

\def\draft{\oddsidemargin -.5truein
        \def\@oddfoot{\sl preliminary draft \hfil
        \rm\thepage\hfil\sl\today\quad\militarytime}
        \let\@evenfoot\@oddfoot \overfullrule 3pt
        \let\label=\draftlabel
        \let\marginnote=\draftmarginnote
   \def\@eqnnum{(\theequation)\rlap{\kern\marginparsep\tt\@eqnlabel}%
\global\let\@eqnlabel\@vacuum}  }

\def\appname{Appendix}
\newcounter{app}
\def\theapp{\Alph{app}}
\def\app{\par
   \addvspace{4ex}
   \@afterindentfalse
  \secdef\@app\@dapp}
\def\@app[#1]#2{\ifnum \c@secnumdepth >\m@ne
        \refstepcounter{app}
        \addcontentsline{toc}{app}{\theapp
        \hspace{1em}#1}\else
      \addcontentsline{toc}{app}{ #1}\fi
   {\parindent \z@ \raggedright
    \Large \bf \appname~\theapp .
   \Large  \bf 
    #2}\nobreak
   \vskip 4ex   \noindent
\setcounter{equation}{0}
\def\theequation{\Alph{app}.\arabic{equation}}}
\def\@dapp#1{%
{\parindent \z@ \raggedright  \bf #1}\par\nobreak}
\def\l@app#1#2{\addpenalty{\@secpenalty}%
   \addvspace{1em plus\p@}%
   \begingroup
   \@tempdima 3em
     \parindent \z@ \rightskip \@pnumwidth
     \parfillskip -\@pnumwidth
     { \bf
     \leavevmode
     #1\hfil \hbox to\@pnumwidth{\hss #2}}\par
     \nobreak
   \endgroup}

\makeatletter
\newdimen\normalarrayskip            
\newdimen\minarrayskip               
\normalarrayskip\baselineskip
\minarrayskip\jot
\newif\ifold             \oldtrue            \def\new{\oldfalse}
\def\arraymode{\ifold\relax\else\displaystyle\fi}
\def\eqnumphantom{\phantom{(\theequation)}} 
\def\@arrayskip{\ifold\baselineskip\z@\lineskip\z@
     \else
     \baselineskip\minarrayskip\lineskip1\baselineskip\fi}

\def\@arrayclassz{\ifcase \@lastchclass \@acolampacol \or
\@ampacol \or \or \or \@addamp \or
   \@acolampacol \or \@firstampfalse \@acol \fi
\edef\@preamble{\@preamble
  \ifcase \@chnum
     \hfil$\relax\arraymode\@sharp$\hfil
     \or $\relax\arraymode\@sharp$\hfil
     \or \hfil$\relax\arraymode\@sharp$\fi}}

\def\@array[#1]#2{\setbox\@arstrutbox=\hbox{\vrule
     height\arraystretch \ht\strutbox
     depth\arraystretch \dp\strutbox
width\z@}\@mkpream{#2}\edef\@preamble{\halign \noexpand\@halignto
\bgroup \tabskip\z@ \@arstrut \@preamble \tabskip\z@ \cr}%
\let\@startpbox\@@startpbox \let\@endpbox\@@endpbox
  \if #1t\vtop \else \if#1b\vbox \else \vcenter \fi\fi
  \bgroup \let\par\relax
  \let\@sharp##\let\protect\relax
  \@arrayskip\@preamble}
\def\eqnarray{\stepcounter{equation}%
              \let\@currentlabel=\theequation
              \global\@eqnswtrue
              \global\@eqcnt\z@
              \tabskip\@centering              
              \let\\=\@eqncr
              $$%
            \halign to \displaywidth  \bgroup
             \eqnumphantom \@eqnsel
      \hskip\@centering                               
    $\displaystyle  \tabskip\z@ {##}$%
    &\global\@eqcnt\@ne \hskip 2\arraycolsep
         $ \displaystyle  \arraymode{##}$\hfil
    &\global\@eqcnt\tw@ \hskip 2\arraycolsep
         $\displaystyle\tabskip\z@{##}$\hfil
         \tabskip\@centering
    &{##}\tabskip\z@\cr}
\makeatother

\newfont{\hr}{msbm10}
\newfont{\ams}{msam10}

\font\numbers=cmss12
\font\upright=cmu10 scaled\magstep1
\def\stroke{\vrule height8pt width0.4pt depth-0.1pt}
\def\topfleck{\vrule height8pt width0.5pt depth-5.9pt}
\def\botfleck{\vrule height2pt width0.5pt depth0.1pt}
\def\Zmath{\vcenter{\hbox{\numbers\rlap{\rlap{Z}\kern 0.8pt\topfleck}\kern
2.2pt
                   \rlap Z\kern 6pt\botfleck\kern 1pt}}}
\def\Qmath{\vcenter{\hbox{\upright\rlap{\rlap{Q}\kern
                   3.8pt\stroke}\phantom{Q}}}}
\def\Nmath{\vcenter{\hbox{\upright\rlap{I}\kern 1.7pt N}}}
\def\Cmath{\vcenter{\hbox{\upright\rlap{\rlap{C}\kern
                   3.8pt\stroke}\phantom{C}}}}
\def\Rmath{\vcenter{\hbox{\upright\rlap{I}\kern 1.7pt R}}}
\def\Z{\ifmmode\Zmath\else$\Zmath$\fi}
\def\Q{\ifmmode\Qmath\else$\Qmath$\fi}
\def\N{\ifmmode\Nmath\else$\Nmath$\fi}
\def\C{\ifmmode\Cmath\else$\Cmath$\fi}
\def\R{\ifmmode\Rmath\else$\Rmath$\fi}

\def\d{\partial}

\def\bea{\begin{eqnarray}}
\def\eea{\end{eqnarray}}

\def\beq{\begin{equation}}
\def\eeq{\end{equation}}
\def\ba{\beq\new\begin{array}{c}}
\def\ea{\end{array}\eeq}
\def\be{\ba}
\def\ee{\ea}
\def\F{{\cal F}}

\def\Q{{\cal Q}}

\def\stackreb#1#2{\mathrel{\mathop{#2}\limits_{#1}}}
\def\Tr{{\rm Tr}}
\def\res{{\rm res}}

\def\half{{\textstyle{1\over2}}}
\def\quot{{\textstyle{1\over4}}}
\def\ha{{1\over 2}}

\def\N2{${\cal N}=2$}
\def\4N{${\cal N}=4$}
\def\1N{${\cal N}=1$}
\def\1N*{${\cal N}=1^*$}

\def\beq{\begin{equation}}
\def\eeq{\end{equation}}
\def\ba{\beq\new\begin{array}{c}}
\def\ea{\end{array}\eeq}
\def\be{\ba}
\def\ee{\ea}
\def\theequation{\thesection.\arabic{equation}}
\def\mezzo#1{\bigskip\noindent{\sl #1}\bigskip}

\newcommand{\rf}[1]{(\ref{#1})}

\begin{document}


\begin{flushright}
FIAN/TD-04/13\\
ITEP/TH-01/13
\end{flushright}
\vspace{1.0 cm}

\begin{center}
\baselineskip20pt
{\bf \LARGE Tau-functions for Quiver Gauge Theories}
\end{center}
\bigskip
\begin{center}
{\large A.~Marshakov}\\
\bigskip
{\em Lebedev Institute, ITEP
and NRU HSE, Moscow, Russia}\\
\medskip
{\sf e-mail:\ mars@lpi.ru, mars@itep.ru}\\
\end{center}
\bigskip

\begin{center}
{\large\bf Abstract} \vspace*{.2cm}
\end{center}

\begin{quotation}
\noindent
The prepotentials for the quiver supersymmetric gauge theories are defined as quasiclassical
tau-functions, depending on two different sets of variables: the
parameters of the UV gauge theory or the bare compexified couplings, and the vacuum condensates of the theory in IR. The bare couplings are introduced as periods on the UV base curve, and
the consistency of corresponding gradient formulas for the tau-functions is proven using the Riemann bilinear relations. It is shown, that dependence of generalised prepotentials for the quiver gauge theories upon the bare couplings turns to coincide with the corresponding formulas for the derivatives of tau-functions for the isomonodromic deformations. Computations for the $SU(2)$ quiver gauge theories with bi- and tri-fundamental matter are performed explicitly and analysed in the context of 4d/2d correspondence.
\end{quotation}

\newpage
\setcounter{equation}0
\section{Introduction}

It has been understood since \cite{gaiotN2dual} that quiver theories with the semisimple gauge groups, consisting of the product of several independent factors, lead
to important new insight on geometry of \N2 supersymmetric gauge theories. In particular they uncover few issues related to the moduli spaces of the base curves, corresponding to gauge theories in the ultra-violet (UV).
Being previously observed only in higher perturbations of the UV prepotential \cite{LNS,GMMMRG,LMN,MN}, this geometry becomes nontrivial for quiver theories and can be studied in detail.

For generic quivers the problem is quite complicated technically, though can be always re-formulated in terms of flat connections on base UV curves \cite{GMN,NRS}. However, it can be described, as usual, in common language of integrable systems \cite{GKMMM}, and this can be
rather instructive for their further study. The important low-energy part of the story for the infra-red (IR) effective theory is contained in the Seiberg-Witten (SW) prepotential \cite{SW},
which is a particular case of the Krichever tau-function \cite{KriW}, if not being restricted to dependence on the IR condensates. The extension to quasiclassical
tau-function becomes necessary to gain the UV information in the final answer as well, e.g. if one
considers a (super)conformal gauge theory. Below we are going to show, that dependence on UV couplings can be considered for the quiver tau-functions almost on equal footing with the dependence upon IR condensates: both are
encoded by the period integrals and satisfy integrability constraints, coming from the Riemann bilinear relations (RBR) on the SW cover $\Sigma$ of the UV curve $\Sigma_0$. The essential difference comes however from the fact, that all ingredients of the tau-function
construction reflecting the UV information can be projected to the base curve $\Sigma_0$, while ``pure IR'' characteristics remain directly related to the form of the cover $\Sigma$ itself.

The starting point for the quiver gauge theory is UV base curve $\Sigma_0=\Sigma_{g_0,n}$ of genus $g_0$ with $n$ punctures, and the complex dimension of its Teichm\"uller space
\be
\label{g0}
L = 3g_0-3+n
\ee
counts the number of the gauge group $G = \bigotimes_{r=1}^L G_r$ factors. For the UV-completeness
each factor should have non-negative beta function, i.e.
\be
\label{beta}
\beta_r = 2N_c^{(r)}-N_f^{(r)}-\sum_{r':\langle r,r'\rangle\neq 0}N_c^{(r')}N_{bf}^{(r,r')} -
\\
-\sum_{r',r'':\langle r,(r',r'')\rangle\neq 0}N_c^{(r')}N_c^{(r'')}N_{3f}^{(r,r',r'')}-\ldots \geq 0,
\ \ \ \ r=1,\ldots,L
\ee
This condition obviously has solutions, when restricted to fundamental and bi-fundamental
matter. The well-known cases are $N_f^{(r)} \leq 2N_c^{(r)}$ with all $N_{bf}^{(r,r')}=
N_{3f}^{(r,r',r'')}=\ldots=0$ (in this case quiver is ``decoupled''), and, for example, conformal quivers with $N_{bf}^{(r,r')}=2$, $N_c^{(r)}=N_c^{(r')}$, $N_{3f}^{(r,r',r'')}=\ldots=0$,
while already for the tri-fundamental matter it can be hardly extended beyond the $SU(2)$ case.
In practice, for getting consistent UV formulation, it restricts the base curves to have $g_0=0,1$, though the higher genera $g_0>1$
can be also considered formally using, for example, the language of pants-decomposition of
$\Sigma_0$ (see, e.g. \cite{Tesch,siqui}). However, the latter class of theories cannot be always formulated in weak-coupling regime \cite{gaiotN2dual,GaMa}
(and presumably does not have generally a Lagrangian description), and this should be reflected in
the problems with their geometric formulation. Fortunately, neither of these complications are present for the $SU(2)$-quivers, when for all gauge group factors $N_c=2$, and even some explicit computations for these theories can be performed and will be presented below.

For rational and elliptic UV curves the geometric picture is rather transparent.
If $g_0=1$ ($n\geq 1$) there exists a canonical holomorphic 1-form $d\omega$ on $\Sigma_0$, e.g. normalized as $\oint_{A^{(0)}}d\omega = 1$ (the base curves $\Sigma_0$ should be always taken with fixed basis of cycles in $H^1(\Sigma_0)$, i.e. over a point in the Teichm\"uller space). For $g_0=0$ one just takes a degeneration of this picture, i.e. considers a sphere with $n\geq 3$ marked points and canonical differential
\be
\label{osphe}
  d\omega = d\Omega^{(0)}_{z_nz_{n-1}} = {dz\over z-z_n}-{dz\over z-z_{n-1}}
  \ee
($z\in \mathbb{C}$, by an $SL(2,\mathbb{C})$-transformation one can always put $z_n=0$ and $z_{n-1}=\infty$). However, there is no such canonical choice for $g_0\geq 2$.

The geometry of the UV curve and its cover can be therefore described in terms of the following co-ordinates:
\begin{itemize}
  \item For $g_0=0$ ($n\geq 3$) fix 3 punctures (e.g. $z_n$, $z_{n-1}$ and $z_{n-2}$), and consider the third kind Abelian differential \rf{osphe}. Then its $(n-3)$ ``periods''
  \be
  \label{crat}
  \log q_j = \int_{B^{(0)}_j} d\omega = \int_{B^{(0)}_j} d\Omega^{(0)}_{z_nz_{n-1}} =\int_{z_{n-2}}^{z_j} d\Omega^{(0)}_{z_nz_{n-1}} =
  \\
  = \log {(z_j-z_n)(z_{n-2}-z_{n-1})\over (z_j-z_{n-1})(z_{n-2}-z_n)},\ \ \ \ j=1,\ldots,n-3
  \ee
  are just the cross-ratios to be identified with the complexified bare couplings, and play the role of desired co-ordinates in what follows.

  For $g_0=1$ ($n\geq 1$) take instead the canonical holomorphic differential $d\omega=dz$, where now $z\in\mathbb{C}/\Gamma(1,\tau_0)$ and consider
  \be
  \label{doper}
  \tau_0 = \oint_{B_0}d\omega = \int_{0}^{\tau_0} dz
  \\
    \tau^{(0)}_j = \int_{B^{(0)}_j}d\omega =\int_{P_n}^{P_j} d\omega = \int_0^{z_j} dz = z_j,\ \ \ \ j=1,\ldots,n-1
 \ee
 with a single fixed puncture $P_n$, which by translation can be always put to $z(P_n)=z_n=0$.
 \item On the cover $\Sigma$, defined by a polynomial equation (with
 the coefficients, taking values in the $k$-differentials on $\Sigma_0$) one
 defines extra period co-ordinates
 \be
 \label{aper0}
 a_I = {1\over 2\pi i}\oint_{A_I} dS$, $I=1,\ldots,g
 \ee
 \end{itemize}

The definition of the Krichever tau-function \cite{KriW} is given by:
 \be
 \label{adper}
 a^D_I = {\d\F\over\d a_I} = {1\over 2\pi i}\oint_{B_I}dS,\ \ \ I=1,\ldots,g
 \ee
 and
 \be
\label{grad}
 {\d\F\over\d\tau^{(0)}_j} = {1\over 2}\int_{A^{(0)}_j} {dS\over d\omega}dS
 \ee
 where $dS$ is the SW differential for the quiver gauge theory. The last formula defines dependence of the tau-function upon the ``extra period'' variables, usually associated with
 the moduli dependent periods of $d\omega$ and leading to appearance of non single valued differentials in the expansion of $dS$. However, when the variables \rf{doper} can be ``projected'' to the base curve and
 associated with degenerate periods (which always happens for $g_0=0$) the r.h.s. of second formula in \rf{grad} is shown below to be reduced to the computation of residues, and its consistency can be proven using the RBR for the second-kind meromorphic Abelian differentials, see Appendix~\ref{ap:RBR}.

The variables \rf{crat}, \rf{doper} can be considered as co-ordinates on the Teichm\"uller space of the base curve $\Sigma_0$. The deformations w.r.t. their higher analogs can be introduced by
     similar to \rf{grad} gradient formulas
  \be
\label{gradhi}
 {\d\F\over\d T^{(k)}_j} = {1\over 2k}\int_{A^{(0)}_j} \left({dS\over d\omega}\right)^{l_k} dS
 \ee
 at least for $T^{(k)}_j=0$, $k>0$ (and $T^{(1)}_j=\tau^{(0)}_j$) with some integers $\{l_k\}$.
 This formula will be discussed below for particular quiver theories.

Let us point out, that the definition of generalised prepotential \rf{adper}, \rf{grad} and
\rf{gradhi} is just a standard definition of quasiclassical tau-function \cite{KriW}, which
has been already partly used in the context of supersymmetric gauge theories. We have to stress however, that formulas \rf{grad} and \rf{gradhi} were never directly applied before in this context, and
only for the quiver gauge theories \cite{gaiotN2dual} their role within the general scheme of constructing the gauge theory prepotentials becomes obvious. It will be also important below
for the particular applications, that gradient formulas \rf{grad}, \rf{gradhi} can be rewritten in terms of residues on the low genera UV curves with punctures, and proof of their consistency in such case is also
essentially simplified.

\setcounter{equation}0
\section{Gauge quivers, complex curves and integrable models}

\subsection{SU(2) quivers and hyperelliptic curves}

For the quiver gauge theory with gauge group $G=SU(2)^{\otimes L}$ coupled to all possible matter
multiplets condition \rf{beta} acquires the form
\be
\label{betasu2}
\beta_r = 4-N_f-2N_{bf}-4N_{3f}-\ldots \geq 0,\ \ \ \ r=1,\ldots,L
\ee
in each $SU(2)$-factor. Such beta-functions vanish and correspond to the superconformal theories for $N_f=2$, $N_{bf}=1$, or
for $N_f=0$, $N_{bf}=2$, when $N_{3f}=\ldots=0$. This happens for UV curves with $g_0=0$ if number of factors is $L=n-3$, and for $g_0=1$ with $L=n$, the corresponding bare couplings are then introduced
by \rf{crat} and \rf{doper} respectively.

These UV variables can be interpreted as co-ordinates on the Teichm\"uller space of the base curve, and to add IR condensates one has to consider the covering curve, or locally - Teichm\"uller deformations. To do this \cite{gaiotN2dual,AGT} one can endow the UV curve $\Sigma_0=\Sigma_{g_0,n}$ with a two-differential
$t=t(z)dz^2$, which at vicinity of each puncture looks as
\be
\label{ts0}
t(z)\ \stackreb{z\to z_j}{=}\ {\Delta_j\over (z-z_j)^2} + {u_j\over z-z_j} + \ldots,\ \ \ \ j=1,\ldots,n
\ee
where the residues $\Delta_j$, $j=1,\ldots,n$ are fixed, and are related to the bare masses of quiver gauge theory. With the fixed residues \rf{ts0} is defined up to a generic holomorphic two-differential
\be
\delta t = \sum_{j=1}^n \delta u_jh_j + \sum_{k=1}^{3g_0-3}\delta y_k h_k
= \sum_{j=1}^n {\delta u_j\over z-z_j} + {\rm reg}
\ee
so totally we have exactly \rf{g0} parameters of deformation, corresponding to a basis in the
space of holomorphic 2-differentials $(\{h_k\},\{h_j\})$,
and in the massless case $\Delta_j=0$ the two-differential \rf{ts0} is holomorphic itself. This case
will be analysed below in detail.

The Seiberg-Witten curve $\Sigma$ in the $SU(2)$-quiver theory covers $\Sigma_0$ twice
\be
\label{SWcu}
x ^2=t(z)
\ee
with branchings at the zeroes $t=0$.
The genus $g$ of $\Sigma$ can be counted, for example, by the Riemann-Hurwitz formula
\be
2-2g = \# S(2-2g_0) - \# BP
\ee
where the number of sheets for the hyperelliptic curve \rf{SWcu} $\# S = 2$, and due to the Riemann-Roch theorem for the 2-differential \rf{ts0} the number of branching points is calculated as
\be
\label{bpRR2}
\# BP = \# ({\rm zeroes}\ t) - \# ({\rm poles}\ t) + 4(g_0-1) = 2n+4g_0-4
\ee
One finds therefore
\be
\label{genRH2}
g = 1 + 2g_0-2+n+2g_0-2 = 4g_0-3+n = L +g_0
\ee
for the full genus of the SW cover $\Sigma$.

\subsection{Gaudin model}

For the lower genera $g_0=0,1$ the curve \rf{SWcu} can be associated with the rational and/or elliptic Gaudin model (see e.g. \cite{Zo,ZoMiMo,nepes}). Consider the meromorphic Lax operator, for example, on sphere $g_0=0$ with $n$ punctures (the elliptic generalisation is straightforward)
\be
\label{Lsl2}
L(z) = \sum_{j=1}^n {A_j\over z-z_j},\ \ \ \ A_j= \left(
\begin{array}{cc}
  \mu_j+h_j & e_j \\
  f_j & \mu_j-h_j
\end{array}\right)\in gl_2
\ee
which defines the following spectral curve equation for the double-cover $\Sigma$ of the rational base
curve $\Sigma_0=\Sigma_{0,n}$
\be
\label{gacu}
\det(x -L(z)) = x ^2 - x \Tr L(z) + \det L(z) = 0
\ee
After the shift $x\to x -\half\Tr L(z) = x - \sum_{i=1}^n{\mu_i\over z-z_i}$, it acquires the form of \rf{SWcu}
\be
\label{gaudincu}
x ^2 = t(z) = - \det L(z)+\quot\left(\Tr L(z)\right)^2  = \half\Tr L(z)^2 - \quot\left(\Tr L(z)\right)^2=
\\
= \sum_{i=1}^n{\half\Tr A_i^2-\quot\left(\Tr A_i\right)^2 \over (z-z_i)^2} + \sum_{i=1}^n{1\over z-z_i}
\sum_{j\neq i}{\Tr (A_iA_j)-\half\Tr A_i\Tr A_j\over z_i-z_j}
\ee
i.e. where
\be
\label{gaudco}
\Delta_j = h_j^2 + e_jf_j = \half\Tr A_i^2-\quot\left(\Tr A_i\right)^2
\\
u_j = \sum_{k\neq j}{2h_jh_k+e_jf_k+e_kf_j\over z_j-z_k}
= \sum_{j\neq i}{\Tr (A_iA_j)-\half\Tr A_i\Tr A_j\over z_i-z_j}
\\
j=1,\ldots,n
\ee
are the Casimir functions (fixed residues) and the $GL(2)$-Gaudin integrals of motion correspondingly, satisfying additional constraints\footnote{In particular, these constraints guarantee the absence of an extra
pole at $z=\infty$, which arises otherwise for the 2-differential $t(z)(dz)^2$ in \rf{gaudincu}, and
therefore ensure correct counting in \rf{bpRR2}, \rf{genRH2}.}
\be
\label{3constr}
\sum_{j=1}^n u_j=0,\ \ \ \ \sum_{j=1}^n (z_ju_j+\Delta_j)
= 0,\ \ \ \ \
\sum_{j=1}^n (z_j^2u_j+2z_j\Delta_j)
= 0
\ee
coming from the transformation properties of the 2-differential in \rf{gaudincu}. Differently, they can be ensured by constraining the total momentum $e=\sum_{j=1}^n e_j=0$ and $f=\sum_{j=1}^n f_j=0$,
and that can be achieved by global $SL(2)$-conjugation of $L(z)$, while $h=\sum_{j=1}^n h_j=0$ corresponds to natural vanishing of the total spin projection.

Equation \rf{gaudincu} exactly corresponds to the double-covering curve \rf{SWcu} of genus $g=n-3$ for the case $g_0=0$. One can now introduce
\be
\label{ldz}
dS = x  dz = {ydz\over\prod_{j=1}^n(z-z_j)}
\ee
where
\be
\label{hypell}
y^2 = R(z) = \sum_{j=1}^n\left(-u_jz+\Delta_j+u_jz_j\right)\prod_{l\neq j}(z-z_l)^2 =
\\
= z^{2n-1}\sum_{j=1}^n u_j + z^{2n-2}\sum_{j=1}^n (z_ju_j+\Delta_j) + z^{2n-3}\sum_{j=1}^n (z_j^2u_j+2z_j\Delta_j)
+ R_{2n-4}(z)
\ee
i.e. upon \rf{3constr} equation \rf{gaudincu} turns into a more common form for a hyperelliptic curve
\be
\label{hypel}
y^2 = R_{2n-4}(z)
\ee
of genus $g=n-3$. It has totally $2n$ punctures, which are pairwise related by hyperelliptic
involution $y\leftrightarrow -y$, and so do the residues of the generating differential $dS = xdz$
\be
\label{masses}
\res_{P_j^\pm} x  dz = \res_{P_j^\pm} {ydz\over\prod_{j=1}^n(z-z_j)} = \pm\sqrt{\Delta_j},
\ \ \ \ j=1,\ldots,n
\ee
which define the mass-parameters in the theory up to their total sum $\sum_{j=1}^n\mu_j$,
absorbed in the shift of $x$-variable in \rf{gaudincu}. In what follows it would be convenient
using $SL(2,\mathbf{C})$-transformations to fix in \rf{ldz} three points to be $(0,1,\infty)$, i.e.
to consider
\be
\label{ldzq}
dS = x  dz = {ydz\over z(z-1)\prod_{j=1}^{n-3}(z-q_j)}
\ee
when one of the points in \rf{hypel} is moved to infinity, giving rise to an odd power polynomial in the r.h.s. Similarly one can treat elliptic case with $g_0=1$, see e.g. \cite{Zo}.

\subsection{SU(N) generalisation}

To complete discussion of the SW curves for quiver theories, let us point out that
the technique of Gaudin models can be also applied to generic quivers, though the results
already for the SW curves are much less practically useful. Apart of the dimensions of corresponding moduli spaces (already derived in the literature, see, for example \cite{NaX,DistlerTach})
we are going however to present some of them here in the form suitable for further study of the tau-functions.

For the $SU(N)$ generalisations, the $gl_N$-valued analog of the Lax operator \rf{Lsl2}
\be
\label{LslN}
L(z) = \sum_{j=1}^n {A_j\over z-z_j} ,\ \ \ \ A_j\in gl_N
\ee
gives rise to the spectral curve equation
\be
\label{NGaudin}
\det(x -L(z)) = x ^N  + \sum_{k=2}^N (-)^kt_k(z)x ^{N-k} = 0
\ee
with the coefficients
\be
\label{2Ndiff}
t_k(z) = \Tr( \underbrace{L(z)\wedge\ldots\wedge L(z)}_k) =
\sum_{j_1\ldots j_k}{\Tr (A_{j_1}\wedge\ldots\wedge A_{j_k})\over(z-z_{j_1})\ldots(z-z_{j_k})} =
\\
= \sum_{j=1}^n\sum_{l=1}^k {u_{l,j}^{(k)}\over(z-z_j)^l},\ \ \ \
k=2,\ldots,N
\ee
being in fact the $k$-differentials on the UV curve $\Sigma_0$.
The total trace can be again absorbed by the shift $x\to x-\Tr L(z) = x-
\sum_{j=1}^N{\mu_j\over z-z_j}$, and all matrices in \rf{2Ndiff} after
redefinition $A_j\to A_j-{\mu_j}\mathbf{1}$ become $sl_N$-valued.

Among the $n\sum_{k=2}^N k = n\left({N(N+1)\over 2}-1\right)$ coefficients in \rf{NGaudin}
there are $n(N-1)$ Casimir functions $\{ u_{k,j}^{(k)}\}=\{ K^{(k)}_j\}$, $j=1,\ldots,n$, $k=2,\ldots,N$, and, taking into account the global group action, one finally gets
\be
\label{imcnt}
n\left({N(N+1)\over 2}-1\right)-n(N-1)-N^2+1 = n{N(N-1)\over 2} - N^2 +1
\ee
integrals of motion of the $SL(N)$ Gaudin model.
This number exactly equals to genus of \rf{NGaudin} in general position: for example,
the Riemann-Hurwitz formula gives now
\be
\label{RHN}
2-2g = \# S(2-2g_0) - \# BP = 2N - \# BP
\ee
where in \rf{NGaudin} the number of sheets of the cover is $\# S = N$, and the number of branching points can be easily found, say, in particular degenerate case $x ^N = t_N(z)$. By the Riemann-Roch theorem, the number of zeroes of the $N$-differential from
\rf{2Ndiff} on sphere is $\#{\rm poles} + 2N(g_0-1) = N(n-2)$ with each such point having multiplicity $(N-1)$, so that
\be
\label{gengau}
\# BP = N(N-1)(n-2)
\\
g = 1-N +\half\# BP = n{N(N-1)\over 2} - N^2 +1
\ee
coinciding finally with \rf{imcnt}. Another way to get the same genus formula
of the SW curve comes from the adjunction formula for the compactified curve \rf{NGaudin}.
Instead of cotangent bundle to the Teichm\"uller space one can consider here the space of the $SL(N,\mathbb{C})$-valued flat connections (see e.g. \cite{NRS} and references therein) on base curve with complex dimension
\be
\label{Nflat}
\dim_\mathbb{C}({\cal M}_{sl_N}(\Sigma_0)) = 2g_0(N^2-1)+n(N^2-N)-2(N^2-1) =
 \\
\stackreb{g_0=0}{=}\ 2\left(n{N(N-1)\over 2}-N^2+1\right) =2g
\ee
One finds that genus \rf{gengau} equals to the half of maximal dimension of the phase space in the $SL(N)$-Gaudin theory. Indeed, formula \rf{Nflat} counts the number of independent parameters in the coefficients of connection \rf{LslN}, subjected to $\sum_{j=1}^nA_j = 0$ (absence of the extra pole in \rf{LslN} at $z=\infty$), with the
fixed traces $\mu_j^{(k)} = \Tr A_j^k$, $k=2,\ldots,N$, and modulo conjugation
by the diagonal gauge group, i.e. for rational curve $\Sigma_0=\Sigma_{0,n}$
\be
\label{dimphasL}
n(N^2-1)-n(N-1)-2(N^2-1) = n(N^2-N)-2(N^2-1) =
\\
= \left.\dim_\mathbb{C}({\cal M}_{sl_N}(\Sigma_0))\right|_{g_0=0,n}
\ee
Such maximal genus and corresponding dimension of the moduli space arise for the theories, associated to the UV curves
having only ``maximal'' punctures with $SU(N)$ flavour symmetry \cite{gaiotN2dual}, e.g. for the $SU(N)^{\otimes (n-3)}$ quiver, coupled with $(n-2)$ copies of $T_N$-theory, however this is not
what one can get from studying the perturbative regime of gauge theory.

For the higher-rank $SU(N)$ quiver theories
one can also have ``special punctures'' \cite{gaiotN2dual} in the
marked points of $\Sigma_{0,n}$. This corresponds to the $SU(r)$ flavour symmetry at each puncture or ``smaller orbits'' described by the rank $r$ matrices $A_j=\textbf{a}^\dagger_j\otimes \textbf{b}_j$ for $j=1,\ldots,n-2$. If $r=1$ one gets the minimal orbits
of dimension $2(N-1)$ ($2N$ components with vanishing $(a^\dagger,b)=0$ taken modulo
$U(1)$ flavour group action). In such case the maximal dimension \rf{dimphasL} is
reduced by $\Delta_{n,N} =
(n-2)(N(N-1)-2(N-1))=(n-2)(N-1)(N-2)$, and becomes
\be
\label{dimphasS}
\left.\dim_\mathbb{C}({\cal M}_{sl_N}(\Sigma_0))\right|_{g_0=0,n} - \Delta_{n,N} =
n(N^2-N)-2(N^2-1) - (n-2)(N-1)(N-2) =
\\
= 2(N-1)(n-3) \equiv 2\tilde{g}
\ee
growing only linearly with $N$. This nicely fits with expression \rf{genRH2} for $g_0=0$, when the dimension of orbit is always as in the $r=1$ case.

For rank $r$ matrices in \rf{LslN} at $j$-th puncture, the maximal order of
the pole
in \rf{2Ndiff} does
not exceed $r$. When all (except for two special marked points, say $z_n$ and $z_{n-1}$) matrices $A_j$ for $j=1,\ldots,n-2$ are of unit rank $A_j\wedge A_j=0$, one gets in \rf{2Ndiff}
\be
u_{k,j}^{(l)} = 0,\ \ \ j=1,\ldots,n-2,\ \ \ l=2,\ldots,k
\ee
Now counting the branching points and the result for smooth genus ${\tilde g}$ of the cover $\tilde{\Sigma}$ is different: instead of \rf{gengau} one gets
\be
\label{genred}
\widetilde{\# BP} = 2(N-1)(n-2)
\\
\tilde{g} = 1-N +\half\widetilde{\# BP} = (N-1)(n-3)
\ee
which exactly fits with \rf{dimphasS}. The number of branch points $\widetilde{\# BP}$, leading to the second line by means of the Riemann-Hurwitz formula, can be computed again when all $k$-differentials vanish in \rf{NGaudin} except for $t_N(z)$. Then all branching
points come with multiplicity $N-1$ for $N$-cover of $\Sigma_0$, and their total number is
$2(n-1)$, coming from the $n-1$ first-order poles at $z_j$, $j=1,\ldots,n-2$ and from $n-1$ zeroes
of the polynomial in numerator
\be
t_N(z) = {Q_{n-1}(z)(dz)^N\over(z-z_n)^N(z-z_{n-1})^N\prod_{j=1}^{n-1}(z-z_j)}
\\
\# ({\rm zeroes}\ t_N) - \# ({\rm poles}\ t_N) = (n-1) - (2N+n-1) = -2N = 2N(g_0-1)
\ee
This analysis shows, that the language of integrable Gaudin models \rf{gacu}, \rf{NGaudin} is useful
for description of
the gauge quivers at least at the level of ``kinematics''. In order to get more information
about the gauge theory one needs to study their prepotentials.

\setcounter{equation}0
\section{Prepotentials and tau-functions}

Let us now turn directly to the tau-functions \cite{KriW} for the $SU(2)$ quiver gauge theories. The SW periods \rf{aper0} are defined for \rf{gaudincu} in a standard way by
\be
\label{aper}
a_I = {1\over 2\pi i}\oint_{A_I} dS = {1\over 2\pi i}\oint_{A_I}x  dz,\ \ \ I=1,\ldots,g
\ee
Additionally, one has to add the variables \rf{crat}, corresponding to the bare couplings. For the rational UV curve $g_0=0$ and fixed $(z_{n-2},z_{n-1},z_n)=(1,\infty,0)$ we introduce
\be
\label{barcp}
i\pi\tau^{(0)}_j = \int_{B^{(0)}_j}{dz\over z} = \int_1^{q_j}{dz\over z} = \log q_j,\ \ \ \
j=1,\ldots,n-3
\ee
and the definition \rf{adper} of the extended prepotential $\F = \F(\mathbf{a};\mathbf{q})$ for quiver theories should be necessarily completed by the gradient formulas \rf{grad}.

The derivatives of so defined tau-functions over the bare couplings
\be
\label{qder}
q_j{\d\F\over\d q_j} = \ha\oint_{A^{(0)}_j} {dS\over d\omega}dS =
\ha\oint_{A^{(0)}_j} x^2 zdz,\ \ \ \ j=1,\ldots,n
\ee
are expressed for the variables \rf{barcp} through the integrals over the dual cycles to $\{B^{(0)}_j\}$ on base curve. Hence, they are expressed through the residues
\be
\label{qderes}
q_j{\d\F\over\d q_j} = \half\res_{q_j} {dS\over d\omega}dS = \half\res_{q_j}x ^2 zdz
= \half(\Delta_j+u_jq_j),\ \ \ \ j=1,\ldots,n-3
\ee
and integrability condition is satisfied, due to
\be
\left.{\d u_i\over\d q_j}\right|_\mathbf{a} = \left.{\d u_j\over\d q_i}\right|_\mathbf{a}
\ee
This becomes a nontrivial relation, when the derivatives are taken at fixed SW periods \rf{aper}, and it turns out to coincide with integrability condition for the tau-functions of the isomonodromic problem, see below.

A direct way to prove the consistency of \rf{qder} in the framework of \cite{KriW} is to use the RBR for the differentials
\be
\label{qdiff}
d\Omega_i = q_i{\d\over\d q_i} dS = q_i{\d x\over\d q_i}\ dz,\ \ \ \ i=1,\ldots,n-3
\ee
However, in contrast to generic situation when the periods are moduli-dependent,
applying such derivatives to \rf{ldzq} gives rise to the \emph{single-valued} Abelian differentials with the second order poles at $z=q_i$
\be
\label{Abel2}
d\Omega_i \stackreb{z\to q_i}{=} {d\xi_i\over\xi_i^2} + \ldots,\ \ \ \ \ \oint_{A_k}d\Omega_i=0,\ \ \ \ i,k=1,\ldots,n-3
\ee
More strictly, in the case of nonvanishing masses \rf{masses} each differential \rf{qdiff} is
in fact a linear combination of several second kind differentials, e.g.
\be
q_i{\d\over\d q_i} dS = d\Omega_i^{(+)}-d\Omega_i^{(-)},\ \ \ \ i=1,\ldots,n-3
\\
d\Omega_i^{\pm} \stackreb{z\to q_i,\pm}{=} {d\xi^\pm_i\over(\xi^\pm_i)^2} + \ldots,\ \ \ \ \ \oint_{A_k}d\Omega_i^{\pm}=0,\ \ \ \ i,k=1,\ldots,n-3
\ee
with the only poles at $z\to q_i$ at each of the sheets of the double cover \rf{SWcu}, \rf{hypel}.
In the massless limit these two copies of poles collide into a single set of the double-poles at the ramification point with $z=q_i$.

Consistency of formulas \rf{qderes} becomes then a consequence of RBR for the second kind meromorphic Abelian differentials (see Appendix~\ref{ap:RBR}), since they can be rewritten in more familiar way
\be
\label{gradres}
q_j{\d\F\over\d q_j} = \half\res_{q_j} {dS\over d\omega}dS =
\half\res_{q_j}\xi_j^{-1}dS,\ \ \ \ j=1,\ldots,n-3
\ee
where the local co-ordinate can be defined by $\xi_j^{-1} = \left.{dS\over d\omega}\right|_{z\to q_j}= \left.xz\right|_{z\to q_j}$ (cf. with \cite{LGGKM,KriW}) in terms of a single meromorphic function ${dS\over d\omega} = xz$. Similarly one can write for the formulas \rf{gradhi} with $k>1$, where
for the $SU(2)$ quivers one has to put $l_k=2k+1$
 \be
\label{gradhires}
 {\d\F\over\d T^{(k)}_j} = {1\over 2k}\res_{q_j} \left({dS\over d\omega}\right)^{2k+1} dS =
 {1\over 2k}\res_{q_j}\xi_j^{-2k-1}dS,\ \ \ \ j=1,\ldots,n-3
 \ee
 which acquire exactly the form of the derivatives of prepotential over the parameters of UV deformation,
 considered in \cite{MN}.

 Consistency of the formulas \rf{adper} follow (due to RBR for the canonical holomorphic differentials,
 see Appendix~\ref{ap:RBR}) from the symmetricity of the period matrix
 \be
 \label{spm}
 T_{IJ} = \oint_{B_I}d\omega_J = { \d a^D_I\over\d a_J} = {\d^2\F\over\d a_I\d a_J}
 \ee
 of the SW curve. Formulas \rf{spm} give the set of low-energy effective couplings, and
 they will be explicitly computed below for certain regions in the moduli space, providing information
 on spectrum of the quiver theory. For the $SU(2)$ quivers this is just an elegant way to reproduce
 what can be read off the week-coupling Lagrangian, but the method itself can be hopefully generalised
 for the nontrivial higher-rank gauge quivers and lead to nontrivial predictions.

\subsection{Massless case}

One can get far more in explicit form, considering the limit of all vanishing flavour masses, when 
\rf{gaudincu} turns into
\be
\label{su2mc}
x^2 = t(z) = {V_{n-4}(z)\over \prod_{i=1}^{n}(z-z_i)}
\\
dS = xdz = \sqrt{V_{n-4}(z)\over \prod_{i=1}^{n}(z-z_i)}dz \rightarrow \sqrt{U}
{\sqrt{\prod_{j=1}^{L-1}(z-v_j)}\ dz\over \sqrt{z(z-1)\prod_{k=1}^L(z-q_k)}}
\ee
where, as usual, it is convenient to fix three of the branching points in the denominator to be $0$, $1$ and $\infty$, and use the number of gauge groups $L=n-3$.
Equation \rf{su2mc} describes itself the \emph{holomorphic} generating differential (notice, that its variations w.r.t. $U$ and $\{ v_j\}$ are also holomorphic and span the $L$-dimensional space of the linearly independent first kind Abelian differentials) on the hyperelliptic curve
\be
\label{Yhol}
Y^2 = z(z-1)\prod_{k=1}^L(z-q_k)\prod_{l=1}^{L-1}(z-v_l)
\ee
of genus $L$. It means, therefore that
\be
\label{holdS}
dS = x  dz = \sum_{J=1}^L a_J d\omega_J
= {dz\over Y}\sqrt{U}\prod_{l=1}^{L-1}(z-v_l)
\\
a^D_I = \oint_{B_I}x dz = \sum_{J=1}^L T_{IJ} a_J
\ee
Computing $q$-derivatives for the differential \rf{holdS} at fixed $a$-periods, one gets
\be
\label{holdSder}
d\Omega_j = q_j{\d\over\d q_j}\ dS = {q_j\sqrt{U}\over 2(z-q_j)}
{\sqrt{\prod_{l=1}^{L-1}(z-v_l)}\ dz\over \sqrt{z(z-1)\prod_{k=1}^L(z-q_k)}}+d\varpi_j,\ \ \ \ \
j=1,\ldots,n-3
\ee
where $d\varpi_j={\phi_j(z)dz\over Y}$ are holomorphic,
fixed by $\oint_{A_i}d\Omega_j=0$. At $z\to q_j$
expressions in \rf{holdSder} behave as
\be
d\Omega_j \simeq {dz \over 2(z-q_j)^{3/2}}
{\sqrt{q_jU\prod_{l=1}^{L-1}(q_j-v_l)} \over \sqrt{(q_j-1)\prod_{k=j}^L(q_j-q_k)}}
\simeq {d\xi_j\over\xi_j^2},\ \ \ \ \
j=1,\ldots,n-3
\ee
where the local co-ordinates now
\be
{1\over\xi_j}  = \left.{dS\over dz/z}\right|_{z\to q_j} = {1\over\sqrt{z-q_j}}{\sqrt{q_jU\prod_{l=1}^{L-1}(q_j-v_l)} \over \sqrt{(q_j-1)\prod_{k=j}^L(q_j-q_k)}},\ \ \ \ \
j=1,\ldots,n-3
\ee
are (conveniently normalised) co-ordinates at ramification points of \rf{Yhol}. We are now going
to discuss these formulas in detail for particular examples.

\subsection{Four flavours and elliptic identities}

In the simplest case of single ($L=n-3=1$) $SU(2)$-gauge theory with four
massless hypermultiplets, with $H=u_2$, $q={(z_2-z_1)(z_4-z_3)\over(z_4-z_2)(z_3-z_1)}$ the differential \rf{ldz} is just
\be
\label{hdS}
dS = x  dz = {\sqrt{H(z_2-z_1)(z_3-z_2)(z_4-z_2)}\ dz\over\sqrt{\prod_{j=1}^4(z-z_j)}}
\rightarrow {\sqrt{Hq(q-1)}\ dz\over\sqrt{z(z-1)(z-q)}}
\ee
the holomorphic differential on torus - a particular case of \rf{holdS} for $L=1$ with $U=Hq(q-1)$.

Its periods can be computed in terms of the elliptic integrals, e.g.
\be
\label{aptor}
a = {1\over 2\pi i}\oint_A x  dz = {\sqrt{Hq(1-q)}\over 2\pi}\oint_A {dz\over\sqrt{z(z-1)(z-q)}} = \sqrt{Hq(1-q)}\ \alpha(\tau)
\ee
where the $A$-cycle encircles $z=0$ and $z=q$. Using the Weierstrass uniformisation with
\be
\label{qthe}
q={e_2-e_3\over e_1-e_3} = {\theta_2^4(0|\tau)\over\theta_3^4(0|\tau)}
\ee
one can directly express
\be
\label{althe}
\alpha(\tau) = {1\over 2\pi}\oint_A {dz\over\sqrt{z(z-1)(z-q)}} = {\sqrt{e_1-e_3}\over\pi}\oint_A {dx\over\sqrt{4(x-e_1)(x-e_2)(x-e_3)}} =
\\
= {2\omega\over\pi}\sqrt{e_1-e_3} = \theta_3^2(0|\tau)
\ee
in terms of the theta-constants. The computation of the dual period on torus obviously gives
\be
\label{bptor}
{\d\F\over\d a} = {1\over 2\pi i}\oint_B x  dz = {\sqrt{Hq(1-q)}\over 2\pi}\oint_B {dz\over\sqrt{z(z-1)(z-q)}} 
\stackreb{\rf{aptor}}{=}\
a\ {\oint_B {dz\over\sqrt{z(z-1)(z-q)}}\over \oint_A {dz\over\sqrt{z(z-1)(z-q)}}} = a\tau
\ee
and, since the modular parameter
of the torus $\tau$ is here independent of $a$, equation \rf{bptor} is trivially solved by $\F = \half\tau a^2$, up to a possible tau-dependent constant. However, it is completed now by the only nontrivial relation for this prepotential, coming from \rf{grad}, i.e.
\be
\label{Ft0}
q{\d\F\over\d q} = \ha\int_{A_0} {dS\over d\log z}dS = \ha\int_{A_0} x ^2 zdz =
\\
= \half Hq(q-1)\int_{A_0}{dz\over(z-1)(z-q)} = {a^2\over 2\alpha^2(\tau)}\int_{A_0}{dz\over(z-1)(z-q)}
\ee
where the last equality is due to \rf{aptor}. The r.h.s. is proportional to $a^2$, and it means that
tau-dependent constant actually vanishes.
The remaining integral is taken over the dual to $B_0$ cycle, defining the only here cross-ratio \rf{crat}
\be
\int_{B_0}{dz\over z} = \int_1^q {dz\over z} = \log q
\ee
or the UV bare coupling $\log q = i\pi\tau_0$. Hence, the $A_0$-period is given by the
residue \rf{gradres}, and therefore \rf{Ft0} gives rise to
\be
{1\over\alpha^2(\tau)}\int_{A_0}{dz\over(z-1)(z-q)} = {1\over\alpha^2(\tau)}\res_{z=q}{dz\over(z-1)(z-q)} =
{1\over\alpha^2(\tau)(q-1)} = {d\tau\over d\tau_0}
\ee
Using \rf{qthe} and \rf{althe}, we therefore obtain
\be
\label{t0t}
{d\tau_0\over d\tau} = \alpha^2(\tau)(q-1) = -{1\over\pi^2}\theta_3^4(0|\tau)\left(1-{\theta_2^4(0|\tau)\over\theta_3^4(0|\tau)}\right) = -{1\over\pi^2}\theta_4^4(0|\tau)
\ee
which is a nontrivial relation between UV and IR couplings $\tau_0$ and $\tau$.
Taking in account the identity\footnote{It can be immediately obtained by taking two derivatives of the addition formula
\be
\label{addthe}
\theta_2(x+y|\tau)\theta_3(x-y|\tau)\theta_2(0|\tau)\theta_3(0|\tau) = \theta_2(x|\tau)\theta_3(x|\tau)\theta_2(y|\tau)\theta_3(y|\tau)-
\theta_1(x|\tau)\theta_4(x|\tau)\theta_1(y|\tau)\theta_4(y|\tau)
\ee
and using $\theta_1'(0|\tau)=\theta_2(0|\tau)\theta_3(0|\tau)\theta_4(0|\tau)$, see e.g.
\cite{japel} for discussion of similar identities.}
\be
\label{id234}
\theta_2''(0|\tau)\theta_3(0|\tau)-\theta_3''(0|\tau)\theta_2(0|\tau) =
- \theta_4(0|\tau)^4\theta_2(0|\tau)\theta_3(0|\tau)
\ee
one finally gets for \rf{t0t}
\be
\label{t0tfin}
i\pi{d\tau_0\over d\tau} =  {1\over\pi i}\theta_4^4(0|\tau) =
-{1\over\pi i}
\left({\theta_2''(0|\tau)\over\theta_2(0|\tau)}-{\theta_3''(0|\tau)\over\theta_3(0|\tau)}\right)=
4{d\over d\tau}\log{\theta_2(0|\tau)\over\theta_3(0|\tau)}
\ee
which is integrated to the Zamolodchikov renormalisation formula $e^{i\pi\tau_0}=q=\theta_2(0|\tau)^4/\theta_3(0|\tau)^4$ \cite{Zam}, see also \cite{ZAGTKl,ZAGT,Pogh}.
This relation was already derived by many different methods, see for example
\cite{Egumar}, where completely different reasoning has been applied, which does not use gradient formulas like \rf{Ft0}. Here it is just the first application of particular case \rf{Ft0} of the general formula \rf{grad}, and we point out, that it
leads to the well-known result only after using not immediately obvious identity \rf{id234}.

In the perturbative limit $q\to 0$ one gets from these formulas
\be
\log q = 4\log {\theta_2(0|\tau)\over\theta_3(0|\tau)} = i\pi\tau + \log 16 + \ldots
\ee
where
\be
\tau_{\rm pert} = {1\over i\pi}\log {(a+m_1)(a+m_2)(a+m_3)(a+m_4)\over (2a)^4}\
\stackreb{m_f=0}{=}\ {1\over i\pi}\log {1\over 16}
\ee
corresponds to finite perturbative renormalisation in massless $N_c=2$, $N_f=4$
superconformal theory.

\subsection{Perturbative computations for bi- and tri- fundamentals
\label{ss:perbitri}}

For a single gauge factor the theory was exactly solved in previous section.
As a next application, consider the case with $L=2$ or $n=5$, which already necessarily contains an external leg, corresponding to the bi-fundamental multiplet. The generating differential \rf{holdS}
\be
\label{hdS2}
dS = x  dz = {\sqrt{U(V-z)}\ dz\over\sqrt{z(z-1)(z-q_1)(z-q_2)}}
\ee
is holomorphic differential on hyperelliptic curve of genus $L=2$, corresponding to (the only
possible for two gauge factors) linear quiver\footnote{Rescaling the variable, e.g. $z=q_2t$, one gets the parameterisation of curves for such quivers used in \cite{gaiotN2dual}.}.

Consider now the perturbative
degeneration, corresponding to the limits $q_1\to 0$, $q_2\to 1$, then \rf{hdS2} can
be explicitly decomposed as
\be
\label{hdS2d}
dS \simeq  {\sqrt{U(V-z)}\ dz\over z(z-1)} = - \sqrt{UV}{\sqrt{V}\ dz\over z\sqrt{V-z}} +
\sqrt{U(V-1)}{\sqrt{V-1}\ dz\over (z-1)\sqrt{V-z}} =
\\
= a_1d\omega_1 + a_2d\omega_2
\ee
where the SW periods are
\be
\label{adS2}
a_1 = {1\over 2\pi i}\oint_{A_1} dS \simeq \res_{z=0} {\sqrt{U(V-z)}\ dz\over z(z-1)} = - \sqrt{UV}
\\
a_2 = {1\over 2\pi i}\oint_{A_2} dS \simeq \res_{z=1} {\sqrt{U(V-z)}\ dz\over z(z-1)} = \sqrt{U(V-1)}
\ee
In this approximation one can substitute into \rf{hdS2}, \rf{hdS2d} expressions for the coefficients of the curve through the periods \rf{adS2}
\be
\label{UVa}
U = a_1^2-a_2^2,\ \ \ \ V={a_1^2\over a_1^2-a_2^2}
\ee
In such limit the elements of the period matrix \rf{spm} are computed the via the integrals
\be
\omega_1 = \int d\omega_1 = \log{\eta-1\over\eta+1},\ \ \ \ \eta^2 = 1-{z\over V}
\\
\omega_2 = \int d\omega_2 = \log{\eta-\sqrt{1-{1\over V}}\over\eta+\sqrt{1-{1\over V}}}
\ee
For example, one can check, that
\be
\label{T2sym}
T_{21} = \oint_{B_2}d\omega_1 \simeq \left. \log{\eta-1\over\eta+1}\right|^{\sqrt{1-{1\over V}}}_{-\sqrt{1-{1\over V}}} = \left. \log{\eta-\sqrt{1-{1\over V}}\over\eta+\sqrt{1-{1\over V}}}\right|^{1}_{-1} \simeq \oint_{B_1}d\omega_2 = T_{12}
\ee
i.e. the period matrix is indeed symmetric, and obtain for
\be
\label{T2diag}
{1\over 2\pi i}T_{11} \simeq {1\over 2\pi i}\left. \log{\eta-1\over\eta+1}\right|^{\sqrt{1-{\epsilon_1\over V}}}_{-\sqrt{1-{\epsilon_1\over V}}}
\simeq {1\over 2\pi i}\log {\epsilon_1^2\over 16V^2} = {1\over i\pi}\log {4\epsilon_1(a_1-a_2)(a_1+a_2)a_1^2\over (2a_1)^4}
\\
{1\over 2\pi i}T_{22} \simeq {1\over 2\pi i}\left. \log{\eta-\sqrt{1-{1\over V}}\over\eta+\sqrt{1-{1\over V}}}\right|^{\sqrt{1-{1+\epsilon_2\over V}}}_{-\sqrt{1-{1+\epsilon_2\over V}}}
\simeq {1\over 2\pi i}\log {\epsilon_2^2\over 16(V-1)^2} =
\\
= {1\over i\pi}\log {4\epsilon_2(a_1-a_2)(a_1+a_2)a_2^2\over (2a_2)^4}
\ee
We have found therefore, that upon identification $4\epsilon_1=q_1$ and $4\epsilon_2=1-q_2$ expressions
\rf{T2diag} reproduce perturbative logarithmic contributions to
the effective couplings of two $SU(2)$ gauge factors with the numerators in the argument of logarithms, coming from fundamental and bi-fundamental matter, and denominators - from the corresponding to each gauge factor vector multiplets. Both factors according to \rf{betasu2} are superconformal gauge theories with
$N_f=2$ and $N_{bf}=1$, since
\be
\label{betaL2}
\beta_r = 4-N_f-2N_{bf} = 0,\ \ \ \ r=1,2
\ee
and we have checked in this way consistency for the first example of nontrivial gauge quivers,
containing bi-fundamental matter.

Similarly one considers the case with $L=3$ $SU(2)$ gauge factors, when the differential \rf{hdS} becomes
\be
\label{hdS3}
dS = x  dz = {\sqrt{U(V_1-z)(V_2-z)}\ dz\over\sqrt{z(z-1)(z-q_1)(z-q_2)(z-q_3)}}
\ee
and we find here a new phenomenon -
in the perturbative limit of such theory with $q_1\approx 0$, $q_2\approx 1$ and $q_3\approx\infty$. One first gets from \rf{hdS3}
\be
\label{adS3}
a_1 = \oint_{A_1} dS \simeq \sqrt{U\over -q_3}\res_{z=0} {\sqrt{(V_1-z)(V_2-z)}\ dz\over z(z-1)} = -\sqrt{U\over -q_3}\sqrt{V_1V_2}
\\
a_2 = \oint_{A_2} dS \simeq \sqrt{U\over -q_3}\res_{z=1} {\sqrt{(V_1-z)(V_2-z)}\ dz\over z(z-1)} = \sqrt{U\over -q_3}\sqrt{(V_1-1)(V_2-1)}
\\
a_3 = \oint_{A_3} dS \simeq \sqrt{U\over -q_3}\res_{z=\infty} {\sqrt{(V_1-z)(V_2-z)}\ dz\over z(z-1)} = -\sqrt{U\over -q_3}
\ee
which corresponds to the expansion of \rf{hdS3}
\be
dS = x  dz \simeq \sqrt{U\over -q_3}{-V_1V_2\ dz\over z\sqrt{(V_1-z)(V_2-z)}} +
\\
+ \sqrt{U\over -q_3}{(V_1-1)(V_2-1)\ dz\over (z-1)\sqrt{(V_1-z)(V_2-z)}}+\sqrt{U\over -q_3}{dz\over \sqrt{(V_1-z)(V_2-z)}}
\ee
over the set of three ``holomorphic'' differentials on degenerate curve of genus $L=3$. Inverting formulas
\rf{adS3}, we find
\be
\label{Va}
V_1V_2 = {a_1^2\over a_3^2},\ \ \ \ \ V_1+V_2 = {a_1^2-a_2^2\over a_3^2} + 1
\ee
and similar to \rf{T2diag} computation gives now
\be
\label{T3diag}
{1\over 2\pi i}T_{33} = {1\over 2\pi i}\oint_{B_3}d\omega_3 \simeq
\\
\simeq {i\over 2\pi}\left.\log\left(z-\half(V_1+V_2)+\sqrt{(z-V_1)(z-V_2)}
\right)\right|^{(\epsilon_3^{-1},+)}_{(\epsilon_3^{-1},-)}
\ee
where the integration limits correspond to the points with $z=\epsilon_3^{-1}$ on ``upper'' and ``lower'' sheets of the double cover. Hence,
\be
\label{T3f}
{1\over 2\pi i}T_{33} \simeq {1\over 2\pi i}\log {\epsilon_3^2(V_1-V_2)^2\over 16} =
\\
= {1\over i\pi}\log {\epsilon_3\sqrt{(a_1+a_3+a_2)(a_1+a_3-a_2)(a_1-a_3+a_2)(a_1-a_3-a_2)}\over 4a_3^4} =
\\
= {1\over i\pi}\log {4\epsilon_3a_3^2\sqrt{(a_1+a_3+a_2)(a_1+a_3-a_2)(a_1-a_3+a_2)(a_1-a_3-a_2)}\over (2a_3)^4}
\ee
where the last expression exactly corresponds to the contribution into the corresponding effective coupling of the vector $SU(2)$ multiplet in denominator under the logarithm, while the numerator together with two fundamental
multiplets contains the contribution of a half-multiplet for the so called \emph{sicilian} quiver. This case again corresponds to the superconformal gauge theory due to \rf{betasu2}, since
\be
\label{betaL3}
\beta_r = 4-N_f-4N_{3f} = 0,\ \ \ \ r=1,2,3
\ee
requires for $N_f=2$, $N_{bf}=0$ fractional value of $N_{3f}=\half$.
The $a$-dependence of perturbative contribution \rf{T3f} for such half-multiplet exactly corresponds to the corresponding structure of 3-vertex in Liouville theory \cite{ZaZa}, this has been already
noticed in \cite{siqui}.
The corresponding perturbative prepotentials
\be
\F_{\rm pert} =
{1\over 2\pi i}\left(\sum_\pm \left(a_1\pm a_2\pm a_3\right)^2\log\left(a_1\pm a_2\pm a_3\right) -
\sum_{i=1,2,3}(2a_i)^2\log (2a_i)\right)
\ee
were already considered in somewhat different, but closely related context in \cite{FeVe}.

In fact, for the differential \rf{hdS2} the calculations are even more straightforward in the perturbative phase with $q_1\to 0$ and $q_2\to\infty$ (instead of $q_2\to 1$). One gets then
\be
dS = \simeq \sqrt{U\over -q_2}{(V-z) dz\over z\sqrt{(z-1)(V-z)}} =
\\
= \sqrt{UV\over q_2}{\sqrt{V}\ dz\over z\sqrt{(1-z)(V-z)}}-\sqrt{U\over q_2}{dz\over \sqrt{(z-1)(z-V)}}
= a_1d\omega_1 + a_2d\omega_2
\ee
with
\be
\label{aUVlim}
a_1 = \res_{z=0} dS = \sqrt{UV\over q_2},\ \ \ \ a_2 = \res_{z=\infty} dS = -\sqrt{U\over q_2}
\ee
and
\be
T_{22} \simeq \oint_{B_2}{dz\over \sqrt{(z-1)(z-V)}} = - \left.\log\left(z-\half(V_1+V_2)+\sqrt{(z-V_1)(z-V_2)}
\right)\right|^{(\epsilon_2^{-1},+)}_{(\epsilon_2^{-1},-)}
\ee
so that
\be
{1\over 2\pi i}T_{22} \simeq {1\over 2\pi i}\log {\epsilon_2^2(V-1)^2\over 16} =
{1\over i\pi }\log {4\epsilon_2a_2^2(a_1-a_2)(a_1+a_2)\over (2a_2)^4}
\ee
In this limit it is also easy and illustrative to compute the
coupling derivatives, for example for $L=2$ we find from \rf{hdS2d}:
\be
\label{L2dS2}
{dS\over d\log z}dS = x^2 z dz = {U(V-z)\ dz\over (z-1)(z-q_1)(z-q_2)}
\ee
so that
\be
\label{L2res}
\res_{z=q_1}{dS\over d\log z}dS = {U(V-q_1)\over (q_1-1)(q_1-q_2)}
\\
\res_{z=q_2}{dS\over d\log z}dS = {U(V-q_2)\over (q_2-1)(q_2-q_1)}
\ee
At $q_1\to 0$, $q_2\to \infty$ it gives after
using \rf{aUVlim}
\be
q_1{\d\F\over \d q_1} \simeq \ha {UV\over q_2} = {a_1^2\over 2}
\\
q_2{\d\F\over \d q_2} \simeq -\ha {U\over q_2} = - {a_2^2\over 2}
\ee
corresponding to small bare couplings $\log q_1 = i\pi\tau_1 \to 0$ and $\log {1\over q_2} =
i\pi\tau_2 \to 0$. In a similar way the proposed formulas can be used for extracting the instantonic
contributions from the geometric data, contained in the equations of the curves.

\setcounter{equation}0
\section{Coupling-derivatives and isomonodromic problem}

Let us now return to the formulas \rf{grad}, and look in detail at their versions \rf{qder}, \rf{qderes}, \rf{gradres},
used above mostly for the UV curves $\Sigma_0=\Sigma_{0,n}$ with particular choice of marked points $(z_1,\ldots,z_n)=(q_1,\ldots,q_{n-3},1,\infty,0)$.
For generic position of the marked points one can write instead of \rf{qderes}
\be
\label{qdegen}
q_j{\d\F\over\d q_j} = \res_{z_j}{dS\over d\omega}dS   = \res_{z_j}\ {x^2 dz\over d\omega/dz} =
\\
= {2z_j-z_n-z_{n-1}\over z_n-z_{n-1}}\Delta_j+{(z_j-z_n)(z_j-z_{n-1})\over z_n-z_{n-1}}u_j
\ \ \ \ j=1,\ldots,n-3
\ee
where the choice \rf{osphe} for $d\omega$ is used. Introduce now
\be
\label{tiso}
{\tilde\F}(\mathbf{q}|\mathbf{a}) = \F(\mathbf{a};\mathbf{q}) - \sum_{j=1}^{n-3}\log q_j^{\Delta_j}
\ee
The last term in the r.h.s. depends only upon the ``unphysical'', non observable in the IR theory, UV
couplings and can be therefore ignored - similarly to the $U(1)$-factors, which often
appear in the context of the AGT-correspondence \cite{AGT}. One can rewrite \rf{qdegen}
for the redefined prepotential \rf{tiso}, with only a little change at the r.h.s.
\be
\label{qdeiso}
q_j{\d{\tilde\F}\over\d q_j} =  2{z_j-z_n\over z_n-z_{n-1}}\Delta_j + {(z_j-z_n)(z_j-z_{n-1})\over z_n-z_{n-1}}u_j,
\ \ \ \ j=1,\ldots,n-3
\ee
Now, change the variables in the l.h.s. of \rf{qdeiso}, according to an obvious rule
\be
\label{chvar}
\tau (z_1,\ldots,z_n|\mathbf{a}) = e^{{\tilde\F} (f(z_1),\ldots,f(z_n)|\mathbf{a})}\prod_{k=1}^n f'(z_k)^{\Delta_k}
\ee
using the fractional-linear M\"obius transformation
\be
\label{flmo}
f(z) = {(z-z_n)(z_{n-2}-z_{n-1})\over (z-z_{n-1})(z_{n-2}-z_n)}
\ee
which is consistent with the choice of the differential \rf{osphe} in the sense, that it is
normalised as
\be
\label{fq}
f(z_n) = 0,\ \ \ f(z_{n-1})=\infty,\ \ \ f(z_{n-2}) = 1
\\
f(z_j) = q_j,
\ \ \ \ j=1,\ldots,n-3
\\
{\tilde\F} (f(z_1),\ldots,f(z_n)) = {\tilde\F} (q_1,\ldots,q_{n-3};1,\infty,0)
\ee
It follows then from \rf{qdeiso} and \rf{chvar}, that
\be
\label{tauu}
{\d\over \d z_j}\log\tau (z_1,\ldots,z_n|\mathbf{a}) = u_j ,\ \ \ \ j=1,\ldots,n
\ee
where $u_j =u_j(z_1,\ldots,z_n|\mathbf{a})$ are the functions of the marked points and the periods. Using \rf{gaudco} for the case of traceless $SL(2)$ gauge connections, one finally rewrites \rf{tauu} in the form
\be
\label{tauiso}
{\d\over \d z_j}\log\tau (z_1,\ldots,z_n|\mathbf{a}) =
\sum_{k\neq j}{\Tr (A_jA_k)\over z_j-z_k},\ \ \ \ j=1,\ldots,n
\ee
which exactly coincides with the corresponding formula for the tau-function
of the isomonodromic problem \cite{SJM}, (see also \cite{GamILis} and references therein, some definitions are collected in Appendix~\ref{ap:isomon}). It is also important to point out, that very similar formulas
with the first-order poles have been discussed for the $c=1$ conformal blocks yet in \cite{ZamAT} (cf.
for example with (3.23) or (3.17) of this paper), and the only difference is that monodromies, playing
the role of the SW periods in Zamolodchikov's case, were independent of positions of all
branch points for the theory of free scalar field.

The detailed discussion of relation of extended quasiclassical tau-function with the tau-function
of the isomonodromic problem goes far beyond the scope of this paper. One can conjecture however,
that it can acquire similar to that of \cite{GamILis} form, since summing up with any
$\mathbf{z}$-independent measure $d\mu(\mathbf{a})$ we find for
\be
\tau (z_1,\ldots,z_n) = \int d\mu({\mathbf{a}})\tau (z_1,\ldots,z_n|\mathbf{a})
\ee
that, following from \rf{tauiso}, one can write
\be
\label{lin}
\overline{u}_j = \overline{u}_j(z_1,\ldots,z_n) =
{\int d\mu({\mathbf{a}})u_j(z_1,\ldots,z_n|\mathbf{a})\tau (z_1,\ldots,z_n|\mathbf{a})
\over \int d\mu({\mathbf{a}})\tau (z_1,\ldots,z_n|\mathbf{a})} =
\\
= {\d\over \d z_j}\log\tau (z_1,\ldots,z_n),\ \ \ \ j=1,\ldots,n
\ee
i.e. the relation \rf{tauiso} is preserved for the generalised prepotentials after taking
any their linear combinations with the $\mathbf{z}$-independent coefficients.

Certainly, this is just a linear relation, which should be treated only as arising in the leading
order of the quasiclassical $\hbar$-expansion, being a particular case of the two-parameter
background deformation of supersymmetric gauge theory, while the nontrivial relation of the formulas \rf{tauiso}, \rf{lin}
with the Painleve-like equations holds for the finite values of $\hbar$. We plan to return
to this point in a separate publication.

\setcounter{equation}0
\section{Discussion}

Let us finally turn to discussion of some general properties of the quiver tau-functions.
First - few remarks about the formulas for ``higher-Teichm\"uller'' deformations \rf{gradhi} and \rf{gradhires}. Similarly to
\rf{L2dS2} and \rf{L2res} one can compute, for example, in perturbative approximation for
the differential \rf{hdS2}:
\be
\label{L2dSk}
\left({dS\over d\log z}\right)^{2k+1}dS = x^{2(k+1)} z^{2k+1} dz =
\\
= U^{k+1}\left({V-z\over (z-1)(z-q_1)(z-q_2)}\right)^{k+1} \ z^k\ dz
\ee
so that
\be
\label{L2resk}
\left.2k{\d\F\over\d T^{(k)}_1}\right|_{T^{(k)}_j=\delta_{k1}\tau^{(0)}_j} =
\res_{z=q_1}\left({dS\over d\log z}\right)^{2k+1}dS\
\stackreb{q_1\to 0}{=}
\\
= \left({UV\over q_2}\right)^k + O(q_1) = a_1^k + O(q_1)
\\
\left.2k{\d\F\over\d T^{(k)}_2}\right|_{T^{(k)}_j=\delta_{k1}\tau^{(0)}_j} =
\res_{z=q_2}\left({dS\over d\log z}\right)^{2k+1}dS\
\stackreb{q_2\to \infty}{=}
\\
= \left({U\over q_2}\right)^k + O\left(q_2^{-1}\right) = a_2^k + O\left(q_2^{-1}\right)
\ee
for all $k>0$. One finds, that formulas \rf{gradhi} and \rf{gradhires} - at least for the case of
$SU(2)$-quivers - give rise to the polynomial deformations of the classical prepotential, describing
theory in UV
\be
\label{UVdef}
\F_{\rm cl} \rightarrow \F_{\rm cl} + \sum_{j,k}T^{(k)}_j {{a_j}^k\over 2k}
\ee
which has been already discussed for a simple gauge group in \cite{LMN,MN}.

Second, we have considered here in detail only the tau-functions for $SU(2)$ gauge quivers. For the
higher rank gauge groups the situation seems to be far more complicated, in particular - when in certain regions
of the moduli spaces only the strong-coupling formulation in terms of the
superconformal theories is known \cite{gaiotN2dual} - for example,
instead of the tri-fundamental matter, considered in sect.~\ref{ss:perbitri}. One can hopefully get more information about such strongly-coupled theories - when studying the tau-functions and
period matrices along the lines proposed in the paper. We have postponed yet the analysis of the higher genus UV curves, but it is especially interesting to get progress for this case, where for higher ranked gauge groups the approach in terms of the gravity duals was developed in \cite{GaMa}.

The extension to the higher-rank gauge theories deserves deeper understanding of the different orbit
structure in the $SL(N)$ Gaudin model at special punctures
and different symplectic leaves in the moduli spaces of $SL(N,\mathbb{C})$ gauge
connections on $\Sigma_0$. Complete analysis of this case requires also the study of
the higher Teichm\"uller spaces and
corresponding deformations of the UV gauge theory, which has been tested above briefly only for the
$SU(2)$ gauge quivers and for the vanishing values of the deformation parameters themselves.
We plan to return to all these questions elsewhere.

\bigskip
\mezzo{Acknowledgements}\\
\noindent
I am grateful to I.~Krichever, J.~Maldacena, A.~Morozov and A.~Rosly
for the very useful discussions and comments, and to the organizers of the meetings in March 2012 in Osaka
and July 2012 in Trieste, when some of these results were presented.
This work was partly supported by RFBR grant 11-01-00962, by joint RFBR project 12-02-92108, by the Program of Support
of Scientific Schools (NSh-3349.2012.2) and by the Russian Ministry of Education under the contract 8207.

\section*{Appendix}
\appendix
\setcounter{equation}0
\section{Riemann bilinear relations
\label{ap:RBR}}

Integrability of the gradient relations for the quasiclassical tau-functions - \rf{adper}, \rf{grad},
\rf{gradhi}, \rf{gradres}, \rf{gradhires} and similar - follow from the Riemann bilinear relations for Abelian differentials \cite{KriW}, which come out of the equality
\be
\label{any}
\int_\Sigma d\Omega_1\wedge d\Omega_2 = 0
\ee
for any two meromorphic differentials $d\Omega_1$ and $d\Omega_2$. The simplest ones, which ensure
consistency of definition for the SW prepotentials \rf{adper}, follow from \rf{any} for any two
canonical first class Abelian or holomorphic
differentials $d\omega_I = {\d\over\d a_I}dS$, $\oint_{A_I}d\omega_J = \delta_{IJ}$, since applying the
Stokes formula on the cut Riemann surface
\be
\label{RBIhol}
0 = \int_{\Sigma_g} d\omega_I\wedge d\omega_J = \int_{\d\Sigma_g} \omega_Id\omega_J
= \sum_{K=1}^g \left(\oint_{A_K}d\omega_I\oint_{B_K}d\omega_J -
\oint_{A_K}d\omega_J\oint_{B_K}d\omega_I\right) =
\\
= T_{IJ} - T_{JI}
\ee
gives immediately the symmetricity of the period matrix \rf{spm}.

For the non-single valued differentials similar arguments work, though derivation requires
more efforts and uses integration by parts, see for example discussion of this issue in
\cite{AMexT}. The non-single valued differentials often arise together with the formulas
of the type \rf{grad}, \rf{gradhi} when taking derivatives over the periods \rf{crat}, \rf{doper}
- as in the case when jumps arise after differentiating a periodic
function over the period. However, after rewriting \rf{gradhi} in the form of \rf{gradres}, the
integrability condition follows already from the RBR for the second-kind Abelian differentials \rf{Abel2}, which are just slight modification of \rf{RBIhol}.

For the meromorphic differentials identity \rf{any} holds on the Riemann surface $\Sigma$ with
punctures, and the boundary of the cut surface $\d\Sigma$ must be supplemented by the contours
$A^{(0)}$, surrounding the punctures (with and their duals $B^{(0)}$). For the differentials \rf{Abel2}
with the second-order poles at $L=n-3$ punctures one gets instead of \rf{RBIhol}
\be
\label{RBImer}
0 = \int_{\Sigma_{g,L}} d\Omega_i\wedge d\Omega_j
= \int_{\Sigma_{g,L}} \Omega_id\Omega_j =
\\
= \sum_{K=1}^g \left(\oint_{A_K}d\Omega_i\oint_{B_K}d\Omega_j -
\oint_{A_K}d\Omega_i\oint_{B_K}d\Omega_j\right) + \sum_{p=0}^L \oint_{A^{(0)}_p}\Omega_id\Omega_j
\ee
The first sum in the r.h.s. vanishes due to canonical normalisation  $\oint_{A_K}d\Omega_i = 0$,
$\forall i$ and $K=1,\ldots,g$, so that the last term gives rise to
\be
\label{intpun}
\oint_{A^{(0)}_i}\Omega_id\Omega_j + \oint_{A^{(0)}_j}\Omega_id\Omega_j = \oint_{A^{(0)}_i}\Omega_id\Omega_j - \oint_{A^{(0)}_j}\Omega_jd\Omega_i = 0
\ee
since $\oint_{A^{(0)}_p}\Omega_id\Omega_j = 0$ for $p\neq i,j$, and in the vicinity of any puncture
Abelian integrals $\Omega_j$ are single-valued so that one can integrate by parts. Each integral in \rf{intpun} can be calculated via the residue
\be
\label{intres}
- \oint_{A^{(0)}_i}\Omega_id\Omega_j = 2\pi i\ \res_{z_i}\xi_i^{-1} d\Omega_j =
2\pi i\ q_j{\d\over\d q_j}\res_{P_i}\xi_i^{-1} dS
\ee
(one can replace $\Omega_j \stackreb{z\to z_j}{\simeq} - {1\over\xi_j} + \ldots$ by its singular part since $d\Omega_j$ are regular at $z=z_j$ for $i\neq j$),
and then equality \rf{intres} ensures consistency of the definition \rf{gradres}.

\setcounter{equation}0
\section{Isomonodromic deformations
\label{ap:isomon}}

We collect here for completeness the formulas defining tau-function of the isomonodromic deformation
problem for the first order $N\times N$ matrix differential equation
\be
\label{sect}
\left({\d\over\d z}-A(z;\{z_j\})\right)\Phi = 0
\\
A(z;\{z_j\}) = \sum_{j=1}^n {A_j\over z-z_j}
\ee
Consider the flat connection in $(n+1)$-dimensional space
\be
\mathcal{A} = Adz + \sum_{j=1}^n \mathcal{A}_j dz_j,\ \ \ \ \ \overline{\mathcal{A}} = 0
\\
d\mathcal{A} + \mathcal{A}\wedge\mathcal{A} = 0
\ee
which satisfies in components
\be
\label{flatcomp}
\d_z\mathcal{A}_i - \d_iA + [A,\mathcal{A}_i]=0
\\
\d_i\mathcal{A}_j-\d_j\mathcal{A}_i + [\mathcal{A}_i,\mathcal{A}_j] = 0
\\
i,j=1,\ldots,n
\ee
For the Schlesinger anzatz
\be
\mathcal{A} = \sum_{j=1}^n {A_j\over z-z_j}d(z-z_j)
\ee
i.e. $\mathcal{A}_j = -{A_j\over z-z_j}dz_j$, the system \rf{flatcomp} turns into
\be
\label{schles}
\d_i A_j + {[A_i,A_j]\over a_i-a_j} = 0,\ \ \ \ i\neq j
\\
\d_iA_i = \sum_{j\neq i}{[A_j,A_i]\over a_j-a_i}
\ee
the system of isomonodromic (by definition!) deformation equations for \rf{sect}. Due to \rf{schles} one can define
\be
d\log\tau = \half\sum_{i\neq j}\Tr(A_iA_j)d\log(z_i-z_j)
\ee
(the one-form in the r.h.s. is closed) the differential of the tau-function, i.e.
\be
{\d\log\tau\over\d z_i} = \sum_{j\neq i}{\Tr(A_iA_j)\over z_i-z_j},\ \ \ i=1,\ldots,n
\ee
and integrability condition for these relations is ensured by \rf{schles}.


\end{document}